\documentclass[aps,prb,superscriptaddress,showpacs,twocolumn]{revtex4}
\usepackage{graphicx}
\usepackage{amssymb}
\usepackage{amsmath}

\newcommand{\bk}{{\bf k}}

\newcommand{\br}{{\bf r}}

\newcommand{\kB}{k_{\mathrm{B}}}

\newcommand{\Tot}{{\mathrm{tot}}}
\newcommand{\SC}{{\mathrm{SC}}}
\newcommand{\FM}{{\mathrm{FM}}}
\newcommand{\SCFM}{{\mathrm{SC\mbox{-}FM}}}

\newcommand{\bbbone}{{\mathchoice {\rm 1\mskip-4mu l} {\rm 1\mskip-4mu l}
          {\rm 1\mskip-4.5mu l} {\rm 1\mskip-5mu l}}}

\newcommand\deleted[1]{{\relax}}

\begin{document}

\title{Pressure dependence of superconducting and magnetic critical
   temperatures\\ in the ruthenocuprates}
\author{R. Citro}
\affiliation{Dipartimento di Fisica ``E. R. Caianiello'', Universit\`a
   di Salerno,\\ and Istituto Nazionale per la Fisica della Materia,
   UdR di Salerno,\\
Via S. Allende, I-84081 Baronissi (SA), Italy}
\author{G. G. N. Angilella}
\affiliation{Dipartimento di Fisica e Astronomia, Universit\`a di
   Catania,\\ and Istituto Nazionale per la Fisica della Materia,
   UdR di Catania, Lab. MATIS,\\ Via S. Sofia, 64, I-95123 Catania, Italy}
\author{M. Marinaro}
\affiliation{Dipartimento di Fisica ``E. R. Caianiello'', Universit\`a
   di Salerno,\\ and Istituto Nazionale per la Fisica della Materia,
   UdR di Salerno,\\
Via S. Allende, I-84081 Baronissi (SA), Italy}
\author{R. Pucci}
\affiliation{Dipartimento di Fisica e Astronomia, Universit\`a di
   Catania,\\ and Istituto Nazionale per la Fisica della Materia,
   UdR di Catania, Lab. MATIS,\\ Via S. Sofia, 64, I-95123 Catania, Italy}

\date{\today}

\begin{abstract}
\medskip
We study the dependence on filling and pressure of the superconducting
   and ferromagnetic critical temperatures of the ruthenocuprates,
   within the two-band model.
At zero pressure, we find separate regions of coexistence of
   superconductivity and ferromagnetism as a function of filling, with
   contiguous regions merging together as pressure increases.
As a function of pressure, a stronger enhancement of the magnetic
   phase results in a reduced pressure effect on the superconducting
   critical temperature.
Comparison with recent experiments on the determination of the
   critical temperatures as a function of the pressure is also
   discussed.
\\
\end{abstract}

\maketitle

\section{Introduction}
\label{sec:intro}

The strong electron correlations in transition metal oxides are known
   to generate complex phase diagrams.
Nonetheless, it was surprising to discover that one such oxide, the
   ruthenocuprate RuSr$_2$GdCu$_2$O$_8$ (Ru-1212), shows a
   low-temperature phase in which homogeneous ferromagnetism and
   superconductivity coexist \cite{Tallon:99,Bernhard:99}.
This new class of materials then adds itself to the metallic
   ferromagnets UGe$_2$ (Ref.~\onlinecite{Saxena:00}), ZrZn$_2$
   (Ref.~\onlinecite{Pfleiderer:01}), and URhGe
   (Ref.~\onlinecite{Aoki:01}), all showing coexistence of
   unconventional superconductivity and itinerant ferromagnetism at
   low temperature ($T\sim 1$~K) and relatively high pressure ($P\sim
   1$~GPa).

The problem of the coexistence of ferromagnetism (FM) and
   superconductivity (SC) is puzzling due to the fact that there are
   at least two factors that would destroy superconductivity in a
   ferromagnetic medium: first, the exchange splitting lifts the
   energy degeneracy of the partners of a spin-up and spin-down Cooper
   pair; second, magnon exchange leads to repulsion for a singlet pair.
Thus, the coexistence of  ferromagnetic order and superconductivity in
   the Ru-1212 and Ru-1222 compounds raises the question how these two
   antagonist states of matter can accommodate each other.
Do both states coexist with no mutual interference or is there a
   competition between superconducting and magnetic order?

Opposite answers come from different experiments.
Recent muon spin rotation experiments on Ru-1212 \cite{Bernhard:99}
   have suggested that the magnetic moments are not affected by the
   appearance of superconductivity below 45~K.
This assumption motivated Shimahara and Hata to study the
   superconducting properties of a system consisting of alternating SC
   and FM layers in the presence of a fixed internal magnetic field,
   neglecting the effect of SC on FM \cite{Shimahara:00}.
On the other hand, experiments on chemical substitution (doping) of
   Ru-1212 indicate that the magnetic and superconducting critical
   temperatures, $T_m$ and $T_c$, respectively, are affected in an
   opposite way with decreasing $T_c$ and an increase of $T_m$ from
   their reference values $T_c \simeq 45$~K and $T_m \simeq 132$~K in
   undoped Ru-1212 \cite{Mandal:02,Klamut:01}.
These results are indicative of a strong competition between
   superconductivity and magnetism.

Since chemical substitution usually affects several parameters at
   the same time, causing changes of the microstructure of the sample, a better
   indication comes from the effect of hydrostatic pressure on the
   superconducting and magnetic phases.
New experiments in hydrostatic pressure on Ru-1212
   \cite{Lorenz:03,Lorenz:03a} show that both $T_c$ and $T_m$ increase
   linearly with pressure $P$, but at different rates.
This rate is distinctly larger for $T_m$ than for the superconducting
   $T_c$, with $dT_c /dP \approx 1$~K/GPa and $dT_m /dP
   \approx 6.7$~K/GPa \cite{Lorenz:03a}.
The relatively small pressure derivative $dT_c /dP$ for the
   superconducting critical temperature is interpreted as an immediate
   consequence of a competition of ferromagnetic and superconducting
   phases: a stronger enhancement of the magnetic phase results in a
   reduced pressure effect on $T_c$ as compared to underdoped
   high-$T_c$ compounds \cite{Lorenz:03a}.

In this paper, we investigate the effect of hydrostatic pressure $P$
   in a phenomenological model of ferromagnetic superconductors
   \cite{Cuoco:03} with two types of carriers pertaining to different
   layers and responsible for superconductivity and ferromagnetism,
   separately.
Our main concern is the dependence of the critical temperatures $T_c$
   and $T_m$ on pressure and filling, that could shed some light on
   the microscopic mechanism of the coexistence or interplay of
   ferromagnetism and superconductivity.
The analysis follows from a detailed study of the dependence of the
   model parameters (hopping and exchange integrals) on pressure.
The paper is organized as follows.
In Sec.~\ref{sec:model} we introduce the microscopic model and discuss
   the phase diagram and the order parameters at finite temperature.
In Sec.~\ref{sec:parma} we give a detailed description of the
   dependence of the model parameters on pressure.
In Sec.~\ref{sec:numerics} we report the numerical results for $T_c$
   and $T_m$.
Finally, in Sec.~\ref{sec:concl} we present our conclusions and give
   directions for future work.

\section{Two-band model}
\label{sec:model}

The unit cell of Ru-1212 may be described as a `double bilayer', each
   bilayer being composed by a CuO$_2$ and a RuO$_2$ layer, separated
   by an intermediate SrO layer.
The two bilayer blocks are in turn separated by a Gd ion, which also
   serves as an inversion point for the unit cell (see, \emph{e.g.,}
   Fig.~1 in Ref.~\onlinecite{Pickett:99}).
As in the high-$T_c$ cuprates, superconductivity is believed to set in
   within the CuO$_2$ layers, while ferromagnetism may be
   thought as mainly due to the ordering of the Ru moments in the
   RuO$_2$ layers.
This has suggested that both the SC and FM phases in the
   ruthenocuprates are not homogeneous at the microscopic scale.
In particular, the SC order parameter may develop \cite{Pickett:99} a
   spatial variation with non-zero total momentum as in the
   Fulde-Ferrell-Larkin-Ovchinnikov (FFLO or LOFF) phase
   \cite{Fulde:64,Larkin:64}, while recent experimental results
   indicate that the FM order is characterized by predominant AFM
   domains separated by nanoscale FM domains \cite{Lorenz:03a}.
A separate origin of SC and FM correlations is also supported by band
   structure calculations \cite{Pickett:99,Nakamura:00}, which
   clearly indicate the existence of electronic subbands pertaining to
   the CuO$_2$ and to the RuO$_2$ layers, respectively, as well as of
   a hybridization term, due to the bridging apical oxygens between
   adjacent layers.
Electronic subbands in the CuO$_2$ and RuO$_2$ layers are mainly
   characterized by the Cu-$3d_{x^2 -y^2}$ and Ru-$4d_{xy}$ orbitals,
   respectively, as well as by the O-$2p_{x,y}$ orbitals
   \cite{Pickett:99,Nakamura:00}.

In order to study the coexistence of SC and FM in ferromagnetic
   metals, such as UGe$_2$ (Ref.~\onlinecite{Saxena:00}), a
   single-band model has been originally developed within the
   mean-field approximation by Karchev \emph{et al.}
   \cite{Karchev:01}, and then numerically discussed by Jackiewicz
   \emph{et al.} \cite{Jackiewicz:03}.
It has been pointed out, however, that a single band model does not
   produce coexistence, but rather a first order transition between
   phases \cite{Joglekar:04}.
On the other hand, coexistence of FM and SC is permitted in multiband
   models \cite{Suhl:01,Abrikosov:01}.

In the case of the ruthenocuprates, a minimal model for coexisting FM
   and SC is then the two-band model of Cuoco \emph{et al.}
   \cite{Cuoco:03}.
There, one may additionally allow for the hybridization of the two
   bands pertaining to the CuO$_2$ and RuO$_2$ layers, respectively,
   by explicitly including an interlayer hopping term.
Accordingly, we assume that the total Hamiltonian can be decomposed as
\begin{equation}
H_\Tot = H_\SC + H_\FM + H_\SCFM ,
\label{eq:ham}
\end{equation}
with $H_\SC$ and $H_\FM$ describing the CuO$_2$ and RuO$_2$ layers,
   respectively, while $H_\SCFM$ contains both the hybridization term
   and exchange correlations between the two subbands.
Within the mean-field approximation, the three terms read in turn:
\begin{subequations}
\label{eq:hams}
\begin{eqnarray}
\label{eq:hamsSC}
H_\SC &=& \sum_{\bk\sigma} \xi_\bk d^\dag_{\bk\sigma} d_{\bk\sigma}
- \sum_\bk [\Delta d^\dag_{\bk\uparrow} d^\dag_{-\bk\downarrow} +
   \mathrm{H.c.}], \\
\label{eq:hamsFM}
H_\FM &=& \sum_{\bk\sigma} \left(\zeta_\bk +\frac{1}{2} \sigma
   M \right) c^\dag_{\bk\sigma} c_{\bk\sigma} ,\\
\label{eq:hamsSCFM}
H_\SCFM &=& \sum_{\bk\sigma} \left[ t_\perp ( d^\dag_{\bk\sigma}
   c_{\bk\sigma} + \mathrm{H.c.} ) \right. \nonumber\\
&&+ \left. \frac{1}{2} \sigma j_\perp M
   d^\dag_{\bk\sigma} d_{\bk\sigma} \right] .
\end{eqnarray}
\end{subequations}
Here, $c^\dag_{\bk\sigma}$, $d^\dag_{\bk\sigma}$ [$c_{\bk\sigma}$,
   $d_{\bk\sigma}$] are creation [annihilation] operators for
   electrons, with wavenumber $\bk$ and spin projection $\sigma=\pm$ or
   $\sigma\in\{\uparrow,\downarrow\}$ along
   a specified direction, in the RuO$_2$ (FM) and CuO$_2$ (SC)
   subbands, respectively;
$\xi_\bk = \epsilon^\SC_\bk -\mu$ and $\zeta_\bk = \epsilon^\FM_\bk
   -\mu$ are the subbands dispersion relations, measured with respect
   to the common chemical potential $\mu$;
$t_\perp$ is the (momentum conserving) interlayer hopping or
   hybridization term between the SC and FM subbands;
$j_\perp = J_\perp /J_\parallel$ denotes the ratio of the interlayer
   to the inplane exchange couplings.
In Eqs.~(\ref{eq:hams}),
\begin{subequations}
\label{eq:OPs}
\begin{eqnarray}
\Delta &=& g \sum_\bk \langle d_{-\bk\downarrow} d_{\bk\uparrow} \rangle ,\\
M &=& J_\parallel \sum_\bk \left( \langle c_{\bk\uparrow}^\dag
   c_{\bk\uparrow} \rangle - \langle c_{-\bk\downarrow}^\dag
   c_{-\bk\downarrow} \rangle \right)
\end{eqnarray}
\end{subequations}
are the mean-field SC and FM order parameters, respectively,
   $\langle\ldots\rangle$ denotes a self-consistent statistical
   average, and $g>0$ is the SC coupling constant, which we assume to
   be independent of momentum, for the sake of simplicity.

Neutron scattering experiments on the magnetic susceptibility
   of the ruthenocuprates \cite{Tallon:99} agree fairly
   well with a MF picture, thus indicating that a MF desription is
   adequate to describe ferromagnetism in the RuO$_2$ planes. 
In order to improve the MF approach, one should take into account for
   the effect of spin density fluctuations via a dynamical
   susceptibility or vertex corrections. 
This improvement has not been considered in our model, where
   the Stoner criterion has been used in order to describe the
   essential aspects of magnetism.

For the band dispersions, within the rigid tight-binding
   approximation, we take
\begin{equation}
\epsilon^{\SC,\FM}_\bk = -2t (\cos k_x + \cos k_y ) + 4t^\prime \cos
   k_x \cos k_y ,
\label{eq:disp}
\end{equation}
where $t$, $t^\prime$ are the appropriate nearest neighbor (NN) and
   next-nearest neighbor (NNN) hopping amplitudes for the two layers,
   respectively.

Apart from a constant term, Eq.~(\ref{eq:ham}) can be conveniently
   rewritten in matrix form as
\begin{equation}
H_\Tot = \sum_\bk B_\bk^\dag \hat{H} B_\bk ,
\label{eq:Htotmat}
\end{equation}
where $B_\bk^\dag = (d_{\bk\uparrow}^\dag ~ d_{-\bk\downarrow} ~
   c_{\bk\uparrow}^\dag ~ c_{-\bk\downarrow} )$ is a four-component
   spinor accounting for the two different orderings,
the real symmetric matrix
\begin{eqnarray}
&&\hat{H} = \nonumber\\
&&\begin{pmatrix}
\xi_\bk + \frac{1}{2} j_\perp M & -\Delta & t_\perp & 0 \\
-\Delta & -\xi_\bk + \frac{1}{2} j_\perp M & 0 & -t_\perp \\
t_\perp & 0 & \zeta_\bk + \frac{1}{2} M & 0 \\
0 & -t_\perp & 0 & -\zeta_\bk + \frac{1}{2} M
\end{pmatrix} \nonumber\\
\label{eq:Hmat}
\end{eqnarray}
has been introduced.
It is worth noting that $\hat{H}$ may be thought as being composed of
   four $2\times 2$ blocks, each diagonal block pertaining to the SC and FM
   subsystems, respectively.
Competition between SC and FM is provided not only by the off-diagonal
   blocks, which only contain the interlayer hopping term $t_\perp$,
   of kinetic origin, but also by the magnetization-induced splitting
   of the SC subband, induced by the interlayer exchange coupling
   $J_\perp$.
Inversely, the presence of superconducting correlations in the CuO$_2$
   layers ($\Delta\neq0$) does not explicitly enter the
   $(2,2)$ FM block, if not, \emph{e.g.,} through the common chemical
   potential $\mu$, to be self-consistently determined as a function
   of the total number of electrons.

Eq.~(\ref{eq:Htotmat}) can be diagonalized by means of standard
   techniques in terms of the four real eigenvalues $E_{\bk\alpha}$
   ($\alpha=1,\ldots 4$) of the band matrix, Eq.~(\ref{eq:Hmat}).
The SC and FM order parameters, $\Delta$ and $M$, can then be
   derived self-consistently from Eqs.~\eqref{eq:OPs} as
   \cite{Leggett:75}:
\begin{subequations}
\label{eq:gapeqs}
\begin{eqnarray}
\label{eq:gapeqSC}
\Delta &=& \frac{g}{4N} \sum_{\bk,\alpha} \frac{\partial
   E_{\bk\alpha}}{\partial
   \Delta} \tanh \left( \frac{\beta E_{\bk\alpha}}{2} \right) ,\\
\label{eq:gapeqFM}
M &=& \frac{J_\parallel}{2N} \sum_{\bk,\alpha} \frac{\partial
   E_{\bk\alpha}}{\partial
   M} \tanh \left( \frac{\beta E_{\bk\alpha}}{2} \right),
\end{eqnarray}
\end{subequations}
at fixed number of electrons $N$ and inverse temperature
$\beta= 1/\kB
   T$.
Numerical analysis of Eqs.~(\ref{eq:gapeqs}) shows \cite{Cuoco:03} that
   the two-band model allows for the coexistence of SC and FM over a
   reasonable range of parameters.
In Eqs.~(\ref{eq:gapeqs}), the eigenenergies $E_{\bk\alpha}$ are
   implicit functions of the order parameters $\Delta$, $M$, via the
   secular equation $\det (\hat{H} - E_{\bk\alpha} \bbbone ) = 0$,
   and their derivatives can be calculated by means of the implicit
   function theorem (Dini's theorem).
By direct inspection of the secular equation, it can be shown that the
   eigenvalues $E_{\bk\alpha}$ are even functions of $\Delta$, while
   there always exist paired branches $\alpha,\bar{\alpha}$ such that
   $E_{\bk\alpha} (-M ) =  -E_{\bk\bar{\alpha}} (M)$.
The critical temperatures $T_c$ and $T_m$ for the onset of SC and FM
   are defined as the largest temperatures for which
   Eq.~(\ref{eq:gapeqSC}) and (\ref{eq:gapeqFM}) have nonzero
   solutions $\Delta$ and $M$, respectively.
They have been obtained by simultaneously solving
   Eqs.~(\ref{eq:gapeqs}), linearized with respect to the appropriate
   order parameter.

\section{Pressure dependence of the model parameters}
\label{sec:parma}

In Eqs.~(\ref{eq:gapeqs}) for the order parameters, pressure $P$ enters
   through the band ($t$, $t^\prime$ in each layers, and $t_\perp$) and
   the coupling parameters ($g$, $J_\parallel$, $J_\perp$), as well as
   through the doping level, here parametrized by the chemical
   potential $\mu$.
The phase diagram of correlated systems close to an ordering
   instability is usually characterized by the interplay of a
   pressure-induced doping variation and any other `intrinsic'
   pressure effect, here accounted for by the pressure-dependence of
   all other model parameters \cite{Schilling:01}.
This scenario can qualitatively explain the pressure dependence of
   $T_c$ in the high-$T_c$ cuprates \cite{Angilella:96}, in particular
   also when an anisotropic doping redistribution takes
   place among inequivalent layers due to an applied pressure
   \cite{Angilella:99b}.
This scenario has been recently related to the proximity of an
   electronic topological transition, where a pressure- or
   strain-induced change of the topology of the Fermi surface takes
   place either because of a change of the electronic structure (at
   constant doping), or because of a doping variation (at fixed or
   rigid band structure, as is usually assumed) \cite{Angilella:02d}.

An estimate of the pressure dependence of the band
   parameters could in principle be achieved through experiments or by
   extensive \emph{ab initio} calculations \cite{Mishonov:00}.
However, due to the limited number of experimental results on
   ruthenocuprates in hydrostatic pressure, in the following we will
   discuss a simplified scheme allowing us to describe the pressure
   variation of the relevant model parameters.

We will be mainly concerned with the pressure dependence of the
    band parameters and of the exchange integrals. Although a pressure
    dependence of the superconducting coupling
   parameter $g$ is also to be expected on general grounds, its actual
   functional form would depend on the microscopic mechanism of
   superconductivity \cite{Schilling:01,Angilella:96}, which is
   currently a matter of debate for the ruthenocuprates.
In view of the reduced pressure effect on $T_c$, as compared to $T_m$,
   we will then neglect altogether the pressure dependence of $g$,
   although, as mentioned in Sec.~\ref{sec:intro}, the enhancement of
   the magnetic phase with increasing pressure could justify a
   reduction of $\partial T_c /\partial P$.

\subsection{Band parameters}

The relevance of the tight-binding approximation for modelling the
   band structure of the CuO$_2$ and the RuO$_2$ layers in both
   cuprate and ruthenate compounds has been reviewed by Mishonov and
   Penev \cite{Mishonov:00}.
A pressure-induced variation of the band parameters entering
   Eq.~(\ref{eq:disp}), namely the nearest-neighbor (NN) and
   next-nearest-neighbor (NNN) hopping amplitudes $t$
   and $t^\prime$, respectively, and of the interlayer hopping
   amplitude, $t_\perp$, entering Eq.~(\ref{eq:Htotmat}), can be
   approximately accounted for within the extended H\"uckel theory
   \cite{Hoffmann:63}.
In this context, such parameters can be roughly approximated by the
   overlap integrals between the appropriate orbitals, which are the
   Cu-$3d_{x^2 - y^2}$ and the O-$2p_{x,y}$, for $t^\SC$;
   the Ru-$4d_{xy}$ and the O-$2p_{x,y}$, for $t^\FM$;
   the O(1)-$2p_{x}$ and O(2)-$2p_{y}$, for $t^\prime$ in both layers; and
   the Ru-$4d_{3z^2 -r^2}$ and O-$2p_z$, for $t_\perp$.
These are two-center integrals, which have been evaluated analytically
   in terms of the distance (and relative orientation) of the two
   orbital centers.
At large intersite distances, the approximate behavior employed
   \emph{e.g.} in Ref.~\onlinecite{Plakida:01} is recovered.

\subsection{Exchange integrals}

In order to calculate the dependence on hydrostatic pressure of the
   exchange interaction in the RuO$_2$ planes, $J_\parallel$, and the
   interlayer exchange coupling $J_\perp$, we follow the approach of
   Munro \cite{Munro:78} using the same approximation scheme and
   generalize it to the case of Ru-$4d_{xy}$ and Cu-$3d_{x^2-y^2}$
   orbitals.
In this approach the quantity $J^{-1} (dJ/dP)$ is determined within
   the theory of solids under hydrostatic pressure in which the
   application of pressure is represented in terms of the crystal
   compressibility and two other parameters associated with the
   electronic screening ($\Lambda$) and the wave-function-distortion
   ($\Omega$). \cite{Munro:78}
To a certain extent, therefore, these two parameters take
   into account for the many-body effects in an ``equivalent
   pressure-free'' model system.

The generic exchange integral at zero pressure is defined by:
\begin{equation}
J=\int d^3 \br_1 d^3 \br_2
\psi^\ast_i (\br_1) \psi^\ast_j (\br_2)
\frac{e^2}{r_{12}}
\phi_i (\br_2) \phi_j (\br_1),
\label{eq:exchint}
\end{equation}
where $r_{12}=|\br_1-\br_2|$, and $\psi_i (\br) = \psi(\br-\br_i )$
   and $\phi_j (\br) = \phi(\br-\br_j )$ are the appropriate
   hydrogenoid orbitals on atoms $i$ and $j$, respectively.
We seek for the pressure dependence of the expression
   (\ref{eq:exchint}) which is to be approximately determined as a
   function of $\Lambda$, $\Omega$ and the compressibility $\kappa$.
To this aim, one first postulates a scaling of the charge-related
   coupling constants which can be written in the forms $e^2
   \rightarrow \Lambda (P) e^2$ and $Z\rightarrow \Omega(P) Z/\Lambda
   (P)$, where $Z$ is the effective charge number of the nuclear
   unit.
Second, one assumes that the fractional variation of the one-electron
   state $\psi_i$ and $\phi_i$ can be written as a function of
   $\kappa$, $\Lambda$, and $\Omega$, \emph{i.e.}
\begin{subequations}
\begin{eqnarray}
\frac{1}{\psi} \frac{\partial \psi}{\partial P} &\approx& f (\kappa,
\Lambda, \Omega), \\
\frac{1}{\phi} \frac{\partial \phi}{\partial P} &\approx& g (\kappa,
\Lambda, \Omega) .
\end{eqnarray}
\end{subequations}
Using the fact that
\begin{equation}
(1/r_{12})^{-1}\ \frac{d(1/r_{12})}{dP} \approx \frac{1}{3}\kappa ,
\end{equation}
we can write
\begin{multline}
\frac{1}{J}\frac{dJ}{dP} \approx \frac{1}{3}\kappa+2f(\kappa, \Lambda,
\Omega)+2g(\kappa, \Lambda, \Omega) +\frac{d \Lambda}{d P} \\
 -\frac{\kappa}{3} \frac{1}{J}
\int d^3 \br_1 \psi^\ast (\br_1 )
\int d^3 \br_2 \psi      (\br_2 )
\frac{e^2}{r_{12}} \\
\times
\left[
\br_j \cdot \nabla \phi^\ast (\br_2 - \br_j ) +
\br_j \cdot \nabla \phi      (\br_1 - \br_j )
\right].
\label{eq:deriv_exch}
\end{multline}
where we have taken $\br_i=0$ and the last term comes from an
   expansion around $\br_j=0$. In evaluating (\ref{eq:deriv_exch}), we
   make use of the following approximations \cite{Munro:78}.
First, it is assumed that the major contribution to the wavefunction
   distortion comes from its radial part, $R(r)$ say, so that
   $\psi^{-1} \partial \psi/ \partial P \approx R^{-1} \partial
   R/\partial P$.
Second, from the same assumption it follows that
\begin{equation}
\label{eq:gradapp} \br_j \cdot \nabla \phi \approx r_j
\frac{\partial \phi}{\partial r} \approx \rho_c \frac{\partial
R}{\partial \rho},
\end{equation}
where $\rho_c=Zc/2 a_0$, $c$ is the distance between the ions and
   $a_0$ is the Bohr radius.

In our specific case, to evaluate $J_\parallel$ we must consider the
   Ru $4d_{xy}$ orbitals, whose radial function, assuming a
   hydrogenoid-like wavefunction, is:
\begin{equation}
R_{4d}(r)= \frac{1}{24\sqrt{10}}\Omega^{7/2} \left(\frac{Z}{2 a_0}
\right)^{3/2}e^{-\Omega
  \rho/2}\rho^2 (6-\Omega \rho),
\label{eq:radial}
\end{equation}
where $\rho=Zr/2 a_0$.
One thus finds
\begin{equation}
\frac{1}{R_{4d}} \frac{\partial R_{4d}}{\partial P}= \frac{1}{2}
\left(\frac{7}{\Omega}-2 \frac{\rho}{6 -\Omega \rho} -\rho\right)
\frac{d\Omega}{dP}. \label{eq:R_deriv}
\end{equation}
In this expression, one makes use of the approximation $\rho \approx
   \langle \rho^2\rangle_{4d}/\langle \rho \rangle_{4d}$ instead of
   the approximation  $\rho \approx \langle \rho\rangle_{4d}$ to
   retain a better numerical accuracy.
Using Eq.~(\ref{eq:gradapp}), we also have:
\begin{equation}
\br_j \cdot \nabla \phi \approx \left( 3
\left\langle\frac{\rho_c}{\rho}\right\rangle_{4d}-\frac{1}{2}
\rho_c \Omega -6 \left\langle\frac{\rho_c}{\rho(6-\Omega
\rho)}\right\rangle_{4d}\right)R_{4d} . \label{eq:grad}
\end{equation}

Setting $f=g$ into Eq.~(\ref{eq:deriv_exch}) and using the relations
   (\ref{eq:R_deriv}--\ref{eq:grad}), we obtain the following
   expression for the variation of the exchange integral $J_\parallel$
   on pressure:
\begin{multline}
\frac{1}{J_\parallel}\frac{dJ_\parallel}{dP} \approx
\frac{1}{3}\kappa \left( 1+\rho_c - 6 \left\langle
\frac{\rho_c}{\rho}\right\rangle_{4d} +12
\left\langle\frac{\rho_c}{\rho(6-\rho)}\right\rangle_{4d}\right)
\\
+\frac{d \Lambda}{dP} + 2\left( 7 -\frac{\langle \rho^2
\rangle_{4d}}{\langle \rho \rangle_{4d}}-2\left\langle
\frac{\rho}{6-\rho}\right\rangle_{4d} \right) \frac{d \Omega}{dP}.
\label{eq:final_expr_jpa}
\end{multline}
where we have expanded up to the lowest contribution in $\Omega$ in
   the coefficients. This expression is equivalent to Eq.~(10) of
   Munro \cite{Munro:78} for the $3d$ orbitals.

In the evaluation of $J_\perp$ we must evaluate the exchange integral
   between the Cu $3d_{x^2-y^2}$  orbital ($\psi$) and the Ru
   $4d_{xy}$ orbital ($\phi$).
The radial function for the $3d_{x^2-y^2}$ orbital is:
\begin{equation}
R_{3d}(r)= \frac{1}{9 \sqrt{30}}\Omega^{7/2} \left(\frac{Z}{2 a_0}
\right)^{3/2}e^{-\Omega
  \rho/2}\rho^2,
\label{eq:radial3d}
\end{equation}
so that we derive:
\begin{equation}
\frac{1}{R_{3d}} \frac{\partial R_{3d}}{\partial P}= \frac{1}{2}
\left(\frac{7}{\Omega}-\frac{\langle \rho^2\rangle_{3d}}{\langle
\rho\rangle_{3d}}\right)\frac{d\Omega}{dP}.
\label{eq:R_deriv_3d}
\end{equation}
Making use of (\ref{eq:gradapp}) and (\ref{eq:radial3d}), we obtain:
\begin{equation}
\br_j \cdot \nabla \phi \approx
\left(2\left\langle\frac{\tilde{\rho}_c}{\rho}\right\rangle_{3d}-\frac{1}{2}
\tilde{\rho}_c \Omega\right)R_{3d} ,
\label{eq:grad3d}
\end{equation}
where $\tilde{\rho}_c=Z\tilde{c}/2 a_0$,  and $\tilde{c}$ is the
   distance between the Ru-Cu ions, this time.
The functions $f$ and $g$ are now:
\begin{eqnarray}
f &=& \frac 1 2\left(\frac{7}{\Omega}-2 \left\langle \frac{\rho}{6
-\Omega \rho} \right\rangle_{4d}- \frac{\langle
\rho^2\rangle_{4d}}{\langle
\rho\rangle_{4d}}\right)\frac{d\Omega}{dP} \nonumber \\
g &=& \frac 1 2\left(\frac{7}{\Omega}- \frac{\langle
\rho^2\rangle_{3d}}{\langle
\rho\rangle_{3d}}\right)\frac{d\Omega}{dP}.
\label{eq:functions}
\end{eqnarray}

Using Eqs.~(\ref{eq:grad3d}) and (\ref{eq:functions}) in
   Eq.~(\ref{eq:deriv_exch}), we obtain the following variation of
   $J_\perp$ with pressure to the lowest order in $\Omega$:
\begin{multline}
\frac{1}{J_\perp}\frac{dJ_\perp}{dP} \approx \frac{1}{3}\kappa
\left( 1+\tilde{\rho}_c - 4 \left\langle
\frac{\tilde{\rho}_c}{\rho}\right\rangle_{3d}\right)+\frac{d
\Lambda}{dP}
\\
+ \left( 7 -\frac{\langle \rho^2 \rangle_{4d}}{\langle \rho
\rangle_{4d}}-\frac{\langle \rho^2 \rangle_{3d}}{\langle \rho
\rangle_{3d}} -4\left\langle \frac{\rho}{6-\rho}\right\rangle_{4d}
\right) \frac{d \Omega}{dP}.
\label{eq:final_expr_jper}
\end{multline}

As a first step, we assume that the contributions from $d \Omega /dP$
   and $d \Lambda /dP$ in Eqs.~(\ref{eq:final_expr_jpa}) and
   (\ref{eq:final_expr_jper}) are small compared to the one from the
   compressibility, and neglect them altogether.
In so doing, the quantities to determine in order to have the full
   dependence of the exchange integrals on pressure, are the
   compressibility $\kappa$, $\rho_c$, $\tilde{\rho}_c$ and all the
   average values that appear in Eqs.~(\ref{eq:final_expr_jpa}) and
   (\ref{eq:final_expr_jper}).

In determining $\rho_c$, we need to know the Ru--Ru ions distance
   $c/a_0$.
The crystal structure analysis of Ru-1212 \cite{Nakamura:00} gives for
   the distance between Ru ions in the RuO layer and the apical oxygen
   of RuO$_6$ octahedra the value $d(\mathrm{Ru-O}_{\mathrm{api}}
   )=1.912$~\AA.
We can then determine the Ru--Ru distance in the layer as $c=4
   d(\mathrm{Ru-O}_{\mathrm{api}} ) \tan \frac{\pi}{6}=4.415$~\AA, or
   $c/a_0= 7.56144$.
Consequently, we obtain $\rho_c\simeq 15.123$.
From the crystal structure analysis we also know the distance between
   Ru and Cu ions,  $d(\mathrm{Ru-Cu})=4.102$~\AA{} and so we obtain
   $\tilde{\rho}_c\simeq 15.508$.
All the average values that appear in Eqs.~(\ref{eq:final_expr_jpa})
   and (\ref{eq:final_expr_jper}) have been evaluated analytically by
   the use of the radial functions Eqs.~(\ref{eq:radial}) and
   (\ref{eq:radial3d}), and their values are reported in
   Table~\ref{tab:averages}.

\begin{table}[tb]
\begin{tabular}{|c|c|}
\hline
$\langle\rho\rangle_{4d}$ = $10.5$ & $\langle\rho\rangle_{3d}$ = $7$\\
$\langle \rho^2 \rangle_{4d}$ = $125.8$ & $\langle \rho^2 \rangle_{3d}$ = $56$ \\
$\displaystyle\left\langle\frac{1}{\rho}\right\rangle_{4d}$ = $0.125$ & $\displaystyle\left\langle\frac{1}{\rho}\right\rangle_{3d}$ = $0.166$\\
$\displaystyle\left\langle\frac{\rho}{6-\rho}\right\rangle_{4d}$ = $-1.75$ & \\
$\displaystyle\left\langle\frac{1}{6-\rho} \right\rangle_{4d}$ =
$-0.125$ & \\
$\displaystyle\left\langle\frac{1}{\rho(6-\rho)}\right\rangle_{4d}$
= $0$ &\\ \hline
\end{tabular}
\caption{%
Average values entering Eq.~(\protect\ref{eq:final_expr_jpa}) and
   Eq.~(\protect\ref{eq:final_expr_jper}), as analytically evaluated
   by means of Eq.~(\protect\ref{eq:radial}) and
   Eq.~(\protect\ref{eq:radial3d}) for the radial functions.}
\label{tab:averages}
\end{table}

Finally, we need the compressibility $\kappa$.
To our knowledge, no experiment has been yet performed to determine
   $\kappa$.
We can estimate this quantity based on the pressure dependence of
   $T_m$ known from experiments \cite{Lorenz:03,Lorenz:03a},
   \emph{viz.} $dT_m/dP=6.7$~K/GPa.
Within Stoner's model of ferromagnetism, one has
$\kB T_m = \pi^{-1} [6(\alpha-1)/\alpha R]^{1/2}$,
where $R= (\rho^\prime /\rho)^2 - \rho^{\prime\prime} /\rho$,
   $\rho=\rho(\mu)$ is the density of states (DOS) at the chemical
   potential $\mu$, and $\alpha=J_\parallel \rho$,
   $\alpha>1$ being the Stoner criterion for ferromagnetism
   \cite{Enz:92}.
Neglecting the pressure dependence of the DOS, one
   roughly finds
\begin{equation}
\frac{1}{T_m} \frac{\partial T_m}{\partial P} = \frac{1}{2}
   \frac{1}{\alpha-1} \frac{1}{J_\parallel} \frac{\partial
   J_\parallel}{\partial P} ,
\end{equation}
which explicitly depends on filling through $\alpha$.
This has to be contrasted with the relation \cite{Munro:78}
\begin{equation}
\frac{1}{T_m}\frac{d T_m}{dP}=\frac{1}{J_\parallel}\frac{d
J_\parallel}{dP},
\end{equation}
holding within the Heisenberg model of ferromagnetism at the
   mean-field level.
Making use of the latter, albeit filling-independent, expression and of
   Eq.~(\ref{eq:final_expr_jpa}), we can estimate
   the compressibility as $\kappa =3.4 \cdot 10^{-3}$~GPa$^{-1}$,
   which is a reasonable value, if compared to the values known for other
   cuprate materials with perovskite structure
   \cite{Cornelius:92,Cornelius:93}. 
Inserting this value back in Eqs.~(\ref{eq:final_expr_jpa}) and
   (\ref{eq:final_expr_jper}), we obtain the dependence of the
   exchange integrals on pressure.
Comparing the relative pressure coefficients, we also obtain $\left(d
   \ln J_\parallel /dP\right)/\left(d \ln J_\perp /dP\right)\sim
   0.77$, indicating that $J_\perp$ increases faster with pressure
   as compared to $J_\parallel$.
We would like to stress that the relation of $T_m$ with the
    relevant model parameters could be different
    if approximations beyond the MF level are taken into
    account.
Neverthless, ac-susceptibility and resistivity measurements on
   ruthenocuprates \cite{Lorenz:03,Lorenz:03a} fairly well agree with
   the MF picture, thus indicating that a MF desription is adequate to
   describe ferromagnetism in the RuO$_2$ planes.

\section{Numerical results}
\label{sec:numerics}

\begin{figure}[t]
\centering
\includegraphics[height=0.9\columnwidth,angle=-90]{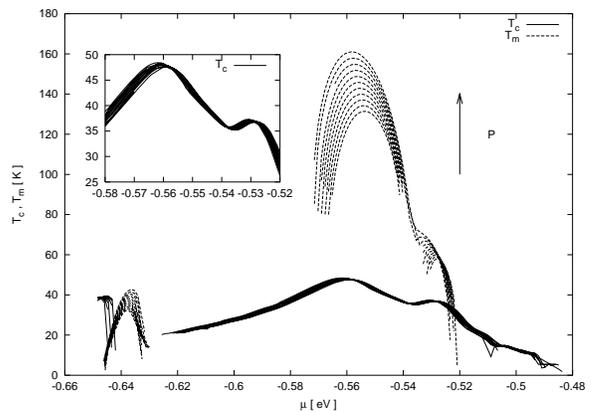}
\caption{Superconducting critical temperature $T_c$ (solid lines,
   and inset) and ferromagnetic critical temperature $T_m$
   (dashed lines), as a function of chemical potential $\mu$, for
   different pressures $P=0-2$~GPa. 
Lower curves correspond to lower pressures, as indicated by the arrow.
}
\label{fig:tpress}
\end{figure}

\begin{figure}[t]
\centering
\includegraphics[height=0.9\columnwidth,angle=-90]{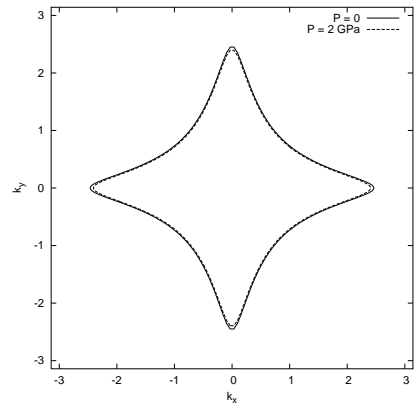}
\caption{Fermi lines of the RuO$_2$ layers corresponding to the
   maximum $T_m$ in Fig.~\protect\ref{fig:tpress}, for $P=0$ and
   $P=2$~GPa.
}
\label{fig:eps}
\end{figure}

In Fig.~\ref{fig:tpress} we show our numerical results for both the
   superconducting critical temperatures, $T_c$, and the ferromagnetic
   critical temperature, $T_m$, as a function of the chemical
   potential $\mu$, for 11 values of pressure $P=0-2$~GPa.
Fig.~\ref{fig:eps} shows the shapes of the Fermi
   surface relative to the RuO$_2$ layer, $\epsilon^\FM_\bk =
   \mu^\ast$, corresponding to the chemical potential $\mu^\ast$ which
   maximizes $T_m$ at a given pressure.
One immediately concludes that, within the present approximation,
   pressure has a negligible effect on the optimal filling for $T_m$.

At zero pressure, we take $t= 0.3$~eV and $t^\prime /t = 0.45$ for
   both the $\SC$ and the $\FM$ bands
   \cite{Pickett:99,Nakamura:00,Kuzmin:00}, and $t_\perp = 0.05t$.
The values of the coupling parameters at zero pressure have been
   chosen so to reproduce the observed optimal values of $T_c$ and
   $T_m$ at $P=0$.
Specifically, we take $g=0.042$~eV, $J_\parallel = 1.4 t$,
   $J_\perp = 0.1 t$.

At zero pressure, we find three separate regions of coexistence of
   superconductivity and ferromagnetism, with $T_m$ displaying three
   pronounced `domes' as a function of chemical potential $\mu$.
This is mainly a consequence of the Stoner criterion, which in the
   simplest version, \emph{i.e.} neglecting interlayer exchange and in
   the absence of competing SC order, reads
   $J_\parallel \rho >1$, so that ferromagnetism is enhanced where the
   DOS is largest, \emph{i.e.} close to Van~Hove
   singularities or electronic topological transitions
   \cite{Angilella:02d}.
As pressure increases, the band widens and the DOS peaks lower
   \cite{Angilella:96}.
However, the exchange couplings $J_\parallel$ and $J_\perp$ are
   expected to increase, as a result of a larger overlap of the
   orbitals in Eq.~\eqref{eq:exchint}, so that the Stoner criterion is
   satisfied over larger filling ranges.
As a consequence, separate regions of coexistence of SC and FM are
   expected to merge, as shown in Fig.~\ref{fig:tpress}.
As is also shown in Fig.~\ref{fig:tpress}, the ferromagnetic
   transition temperature is found to increase with pressure at a rather
   large rate, in good qualitative agreement with experiments.
Its values go
   from $130$~K  to $160$~K in the pressure range $P=0-2$~GPa.
Thus the ferromagnetic state appears to be strongly stabilized
   under pressure which should have some consequences for the
   superconducting state.

As mentioned in Sec.~\ref{sec:parma}, up to now we have neglected any
   explicit pressure effect on the SC coupling parameter $g$.
This is motivated by a lack of either theoretical or phenomenological
   input for the microscopic mechanism of superconductivity in the
   ruthenocuprates, which is expected to be of unconventional nature,
   as is possibly for the high-$T_c$ cuprates.
Quantitatively, this approximation should not affect much our results,
   in view of the small pressure effect on $T_c$, as compared to
   $T_m$.
However, we have numerically studied the competition of SC and FM,
   both at zero pressure and for increasing $P$, by tentatively
   assuming a small linear dependence of $g$ on pressure.
Indeed, the effect of a competing FM phase at $P=0$ does decrease
   $T_c$, compared to the case $J_\parallel = J_\perp = 0$ case, as
   already observed by Cuoco \emph{et al.} \cite{Cuoco:03}.
This tendency is also confirmed at nonzero pressure.

Since we neglected any explicit dependence of the SC coupling constant
   $g$ on pressure, the albeit small increase of $T_c$ shown in the
   inset of Fig.~\ref{fig:tpress} must be mainly attributed to the
   pressure-induced changes of the kinetic terms in the SC and SC$+$FM
   Hamiltonians, Eqs.~(\ref{eq:hamsSC}) and (\ref{eq:hamsSCFM}),
   \emph{i.e.} changes in the band structure.
Although the fine details of the variations of $T_c$ and $T_m$ are
   related to each other in an inherently nonlinear way through
   Eqs.~(\ref{eq:gapeqs}), one expects that a pressure-induced
   enhancement of the hopping parameters $t$, $t^\prime$, and
   $t_\perp$ results in a shift towards the band bottom of the
   Van~Hove singularity pertaining to the SC subband, accompanied by a
   steepening of the DOS, which is indeed recovered in the tendency of
   the maxima in the $T_c$ curves to move towards lower chemical
   potentials with increasing pressure (Fig.~\ref{fig:tpress}, inset).

\section{Conclusions}
\label{sec:concl}

We have considered the two-band model for the coexistence of
   superconductivity and ferromagnetism in the ruthenocuprates.
We have self-consistently solved the equations for the SC
   (respectively, FM) critical temperature in the presence of FM
   (respectively, SC) order, both as a function of filling (here
   parametrized by the chemical potential $\mu$), and as a function of
   pressure.
We find separate filling ranges where the coexistence of SC and FM is
   allowed, merging into larger ranges, as the Stoner criterion gets
   more effective with increasing pressure.
The ferromagnetic transition temperature is found to increase
   with pressure at a rate distinctly larger than that of the
   superconducting temperature, in good qualitative agreement with
   recent experiments in the ruthenocuprates.
Due to the competition
   between superconductivity and ferromagnetism, the stronger
   enhancement of the magnetic phase results in a suppression of
   the pressure effect on $T_c$.

Our present model does not account for spatial variations of both
   order parameters, or of their symmetry in $\bk$-space, which has
   been assumed to be $s$-wave, for the sake of simplicity.
In particular, a $c$-axis modulation of the order parameters may be
   important in view of the stronger pressure dependence of the
   interlayer correlations.
A more detailed study of the pressure dependence of the critical
   temperatures would require more reliable estimates of the compressibility
   and of the pressure dependence of the charge filling, which await
   more experimental work.

\begin{acknowledgments}
We thank M. Cuoco, G. Forte, P. Gentile, C. Noce, F. Siringo, and
   M. Capriolo for useful discussions on the two-band model.
We also thank F. Mancini for his continued interest and encouraging
   support in our work.
\end{acknowledgments}

\begin{small}
\bibliographystyle{apsrev}
\bibliography{a,b,c,d,e,f,g,h,i,j,k,l,m,n,o,p,q,r,s,t,u,v,w,x,y,z,zzproceedings,Angilella}
\end{small}

\end{document}